\documentclass[twocolumn,eqsecnum,floatfix,aps,prc]{revtex4}
\usepackage{epsfig}
\usepackage{times}

\def\lap{\mathrel{\mathpalette\fun <}}

\def\fun#1#2{\lower3.6pt\vbox{\baselineskip0pt\lineskip.9pt
  \ialign{$\mathsurround=0pt#1\hfil##\hfil$\crcr#2\crcr\sim\crcr}}}
\begin{document}


\title { Linear correlations between 
$^4$H\lowercase{e} trimer and tetramer energies \\
calculated with various realistic $^4$He potentials 
}

\author{E.\ Hiyama}
\email{hiyama@riken.jp}
\affiliation{RIKEN Nishina Center, RIKEN, Wako 351-0198, Japan}

\author{M.\ Kamimura}
\email{mkamimura@riken.jp}
\affiliation{Department of Physics, Kyushu University,
Fukuoka 812-8581, Japan, \\
and RIKEN Nishina Center, RIKEN, Wako 351-0198, Japan}

\date{\today}

\begin{abstract}
In a previous work [Phys. Rev. A {\bf 85}, 022502 (2012)]
we calculated, with the use of the 
Gaussian expansion method for few-body systems, 
the energy levels and spatial structures of
the $^4$He trimer and tetramer ground and excited states using
the LM2M2 potential, which has a very strong short-range repulsion. 
In this work, we calculate the same quantities
using the current most accurate $^4$He-$^4$He potential 
[M. Przybytek {\it et al.}, Phys. Rev. Lett.  {\bf 104}, 183003 (2010)]
that includes the adiabatic, relativistic, QED and residual 
retardation corrections.
Contributions of the corrections to the 
tetramer ground-(excited-)state energy $-573.90\,(-132.70)$ mK 
are respectively 
$-4.13\,(-1.52)$ mK, $+9.37 \,(+3.48)$ mK,   $-1.20 \,(-0.46)$ mK and
\mbox{$+0.16\,(+0.07)$ mK}.
Further including other realistic $^4$He potentials, 
we calculated the binding energies of  the trimer and tetramer
ground and excited states, 
$B_3^{(0)}, B_3^{(1)}, B_4^{(0)}$ and $B_4^{(1)}$, respectively.
We found that the four kinds of the binding energies
for the different potentials 
exhibit perfect linear correlations 
between any two of them
over the range of binding energies relevant for $^4$He atoms
(namely, six types of the generalized Tjon lines are observed). 
The dimerlike-pair model for $^4$He clusters, proposed in the previous work, 
predicts a simple interaction-independent relation
$\frac{B_4^{(1)}}{B_2}=\frac{B_3^{(0)}}{B_2} + \frac{2}{3}$, which
precisely explains the \mbox{correlation}  between 
the  tetramer excited-state energy and
the trimer ground-state energy, with $B_2$ being the
dimer binding energy.
\end{abstract}

   
\maketitle

\section{INTRODUCTION}

The bosonic $J=0^+$ three and four $^4$He \mbox{atom systems},
which are very weakly bound under the $^4$He-$^4$He potential with
an extremely strong repulsive core
followed by the van der Waals attraction,  
are known to be suitable for studying the 
Efimov effect  and the universality 
in the systems interacting with large
scattering length~\cite{Efimov70,Efimov11,Braaten06,Braaten03,Platter04}. 

In a previous paper~\cite{Hiyama2012}, referred to as paper I in the following, 
we presented  state-of-the-art four-body calculations for
the $^4$He tetramer ground- and excited-state binding energies 
and structural properties  using a realistic $^4$He potential
called LM2M2~\cite{LM2M2}, which has a very strong short-range repulsion. 
At the same time, our three-body calculation
reproduced all the well known results for the $^4$He trimer.
We took the Gaussian expansion method (GEM) for \mbox{{\it ab initio}}
variational calculations of few-body 
systems~\cite{Kamimura88,Kameyama89,Hiyama03}.
The total wave function is expanded in terms of
totally symmetrized few-body Gaussian basis functions,
ranging from very compact to very diffuse with the Gaussian ranges 
in geometric sequences.

The method is suitable for describing
the short-range correlations 
(without \mbox{{\it a priori}} assumption of any two-body 
correlation function) and 
the long-range asymptotic behavior 
(see the review papers~\cite{Hiyama03,Hiyama09,Hiyama10,Hiyama12FB} for
many applications of the GEM).
As a result, we found in paper I that 
precisely the same shapes of the short-range 
correlation ($ r_{ij} \lap 4 $\AA) in  the dimer appear in 
the ground and excited states of the trimer and tetramer and that 
the wave functions
of the very weakly-bound excited states of
the trimer and the tetramer reproduce
the correct asymptotic behavior up to
up to $\sim\!1000 $\AA.

Recently, Przybytek {\it et al.}~\cite{PCKLJS} proposed 
a $^4$He pair potential that is currently most accurate.
Such an accurate $^4$He potential is of importance,
according to Ref.~\cite{PCKLJS},
in several branches of science, for example, 
in metrology (thermodynamics standards)~\cite{metrology1,
metrology2,metrology3},
helium-nanodroplet spectroscopy~\cite{spectro1,spectro2}, 
and low-temperature condensed matter 
physics~\cite{condens} as well as in the study of the
unusually large and very weakly bound states of the $^4$He clusters.
The potential of Ref.~\cite{PCKLJS} 
includes, in addition to the standard 
Born-Oppenheimer (BO) potential, 
various post-BO contributions. 
 The main contributions are
(i) the adiabatic corrections 
resulting from the leading-order coupling of the electronic 
and nuclear motions, (ii) the relativistic corrections 
to the Schr\"{o}dinger equation, (iii) the
quantum electrodynamics (QED) corrections, and
(iv) the residual retardation correction.
The largest contribution to
the dimer energy ($-1.615$ mK) is $+0.226$ mK repulsively 
from the correction (ii), and the total contribution 
is $+0.103$ mK with a mutual cancellation among (i)--(iv)~\cite{PCKLJS}.
The potential is referred to as ``PCKLJS'' (an acronym for
Przybytek-Cencek-Komasa-Lach-Jeziorski-Szalewicz, the authors of Ref.
~\cite{PCKLJS}) in a subsequent 
paper~\cite{Spirko}; hereafter we  use this acronym.

The first purpose of the present work is to calculate,
using the potential PCKLJS, 
the binding energies of  the trimer and tetramer
ground and excited states, $B_3^{(0)}, B_3^{(1)}, B_4^{(0)}$ 
and $B_4^{(1)}$, respectively,
together with the estimation of the individual contributions
of the corrections (i)--(iv).

 The large scattering length of $^4$He-$^4$He potential
leads to universal properties in the four-body problem.
An example is  existence of the correlations between 
the different observables. Thus,
the second purpose of this work is 
to calculate the binding energies $B_3^{(0)}, B_3^{(1)}, B_4^{(0)}$ 
and $B_4^{(1)}$ 
using various realistic $^4$He potentials
and investigate six types of  the correlations between any two of 
the four energies. 
The potentials employed are PCKLJS and  
six other potentials: 
LM2M2~\cite{LM2M2},TTY~\cite{TTY},  
HFD-B~\cite{HFD-B}, HFD-B3-FCI1~\cite{B3FCI1a,B3FCI1b,SAPT2},
SAPT96~\cite{SAPT96a,SAPT96b,SAPT2} 
and \mbox{CCSAPT07}~\cite{CCSAPT07}
(see Ref.~\cite{Szalewicz-review} for a review of
the recent study of the $^4$He potential); in the last three,
we choose the cases of the retardation corrections
included.

Recently, universal correlations between observables have been 
studied extensively in four-boson systems interacting
with large scattering length~\cite{Platter04,Hammer07,Stecher09,
Stecher09PRL,Tomio06,Tomio2011PRL,Tomio2011FBS}.
As for the specific $^4$He tetramers,
the universal scaling functions
for the correlations between the trimer and tetramer binding energies
were obtained by the 
leading-order effective theory~\cite{Braaten03,Platter04}
and compared with the energies calculated using
realistic $^4$He potentials.
However, due to the scarce calculation of the
$^4$He tetramer excited-state binding energy $B_4^{(1)}$ 
using the realistic pair potential
at that time, the correlations associated with $B_4^{(1)}$ 
remained unexplored (see Fig.~4 of Ref.~\cite{Platter04}). 
In the present work, we provide
precise systematic results on the six types of correlations and
demonstrate that the correlations are all linear
over the range of binding energies relevant to $^4$He atoms.
We compare the result  with that given by the 
leading-order effective theory~\cite{Braaten03,Platter04}
for $^4$He atoms.


This paper is organized as follows:  In Sec.~II,
we briefly present our calculational method 
GEM~\cite{Kamimura88,Kameyama89,Hiyama03}.
Calculated results using the PCKLJS potential
are presented in Sec.~III
together with those for the post-BO corrections (i)--(iv).
In Sec.~IV, using various $^4$He potentials, we calculate
the trimer and tetramer ground- and excited-state
binding energies and discuss the correlations between them
in comparison with the universal scaling functions obtained
by the leading-order effective theory for the $^4$He atom.
A summary is given in Sec.~V.

\section{Method}

We employ the same  {\it ab initio}  variational 
method GEM as in the previous 
work~\cite{Hiyama2012} to solve the  
ground and excited states of the $^4$He trimer and tetramer.
Here, we review the method 
for the case of the tetramer.

\begin{figure}[h]
\begin{center}
\epsfig{file=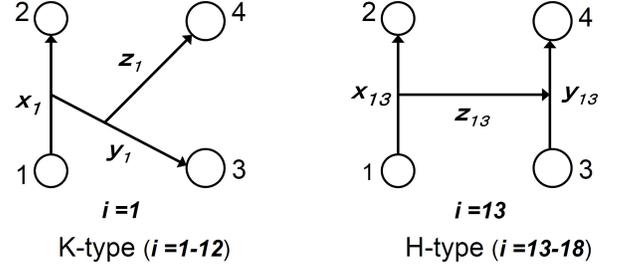,scale=0.180}  
\end{center}
\caption{ $K$-type and $H$-type Jacobi coordinates for the
$^4$He tetramer. Symmetrization of the four particles 
generates the sets $i=1, ..., 12$ ($K$-type) and 
$i=13, ...,  18$ ($H$-type).
}
\label{fig:4bodyjacobi}
\end{figure}

We take two types of Jacobi coordinate sets, $K$-type and $H$-type
(Fig.~\ref{fig:4bodyjacobi}).
For the $K$-type, ${\bf x}_1={\bf r}_2-{\bf r}_1$,
${\bf y}_{1}={\bf r}_3 -\frac{1}{2}({\bf r}_1 + {\bf r}_2)$ and
${\bf z}_{1}={\bf r}_4 -\frac{1}{3}({\bf r}_1 +{\bf r}_2 + {\bf r}_3)$ 
and cyclically for $\{{\bf x}_i, {\bf y}_i, {\bf z}_i;\, i=2, ..., 12\}$ 
by the symmetrization between the four particles.
For the $H$-type,
${\bf x}_{13}={\bf r}_2-{\bf r}_1$,
${\bf y}_{13}={\bf r}_4-{\bf r}_3$, and
${\bf z}_{13}=\frac{1}{2}({\bf r}_3 + {\bf r}_4)
 -\frac{1}{2}({\bf r}_1 + {\bf r}_2)$ 
and cyclically for $\{{\bf x}_i, {\bf y}_i, {\bf z}_i;\, i=14, ..., 18\}$. 

The total four-body  wave function $\Psi_4$ is
to be obtained by solving the Sch\"{o}dinger equation
\begin{equation}
( H - E ) \Psi_4 =0
\end{equation}
with the Hamiltonian 
\begin{equation}
H= -\frac{\hbar^2}{2\mu_x} \nabla^2_x
   -\frac{\hbar^2}{2\mu_y} \nabla^2_y
   -\frac{\hbar^2}{2\mu_z} \nabla^2_z
 + \sum_{1=i<j}^4 V(r_{ij}),
\label{eq:Hamil4}
\end{equation}
where $\mu_x=\frac{1}{2} m$, $\mu_y= \frac{2}{3} m$ and 
$\mu_z= \frac{3}{4} m$ on the $K$-type coordinates, and
$\mu_x=\mu_y=\frac{1}{2} m$ and $\mu_z=m$ on the $H$-type ones, 
$m$ being the mass of the $^4$He atom.  
We take $\frac{\hbar^2}{m}=12.11928$ K\AA$^2$~\cite{newmass}.
$V(r_{ij})$ is the two-body $^4$He-$^4$He potential
as a function of the pair separation 
${\bf r}_{ij}={\bf r}_j- {\bf r}_i$.

The wave function $\Psi_4$ is
expanded in terms of the  symmetrized 
$L^2$-integrable $K$-type and $H$-type four-body
basis  functions:
\begin{equation}
 \Psi_4 =
\sum_{{\alpha_{K}}=1}^{\alpha_{K}^{\rm max}}
  A^{({K})}_{\alpha_{K}} 
\Phi^{({\rm sym};K)}_{\alpha_{K}} +
\sum_{{\alpha_{H}}=1}^{\alpha_{H}^{\rm max}}
  A^{({H})}_{\alpha_{H}} \Phi^{({\rm sym};H)}_{\alpha_{H}} ,
\end{equation}
\vskip -0.3cm
\begin{eqnarray}
 \Phi_{\alpha_{K}}^{({\rm sym};K)} &=& \sum_{i=1}^{12} 
    \Phi^{({K})}_{\alpha_{K}}
( {\bf x}_i, {\bf y}_i, {\bf z}_i ), \\
 \Phi_{\alpha_{H}}^{({\rm sym};H)} &=& \sum_{i=13}^{18} 
    \Phi^{({H})}_{\alpha_{H}}
( {\bf x}_i, {\bf y}_i, {\bf z}_i ), 
\end{eqnarray}
in which $({\bf x}_i, {\bf y}_i, {\bf z}_i )$ is the 
$i$-th set of Jacobi coordinates.
It is of importance that $ \Phi_{\alpha_{K}}^{({\rm sym};K)} $
and $ \Phi_{\alpha_{H}}^{({\rm sym};H)} $
are constructed on the full 18 sets of
Jacobi coordinates; this makes the function space of the basis 
quite wide.

The eigenenergies $E$ and amplitudes 
$A_{\alpha_{K}}^{(K)}  (A_{\alpha_{H}}^{(H)}) $ 
are determined
by the Rayleigh-Ritz variational principle:
\begin{eqnarray}
&&\langle \: \Phi_{\alpha_{\rm K}}^{({\rm sym;K})}
 \:| \:H - E \: |\: \Psi_4 \: \rangle =0, \\
&& \langle \: \Phi_{\alpha_{\rm H}}^{({\rm sym;H})}
 \:| \:H - E \: |\: \Psi_4 \: \rangle =0, 
\end{eqnarray}
where $\alpha_{\rm K}=1, ...,  \alpha_{\rm K}^{\rm max}$
and   $\alpha_{\rm H}=1, ...,  \alpha_{\rm H}^{\rm max}$.
These equations result in the generalized matrix eigenvalue problem
[Eqs.(3.8)-(3.10) of paper I].

We describe the basis function
$\Phi^{({K})}_{\alpha_{K}}(\Phi^{({H})}_{\alpha_{H}})$ 
in the form
\begin{eqnarray}
 \Phi^{({K})}_{\alpha_{K}}( {\bf x}_i, {\bf y}_i, {\bf z}_i ) 
= \phi^{(^{\rm cos}_{\rm sin})}_{n_x l_x}(x_i) \,
  \phi_{n_y l_y}(y_i) \,
  \varphi_{n_z l_z}(z_i) \nonumber \\
\times \Big[ \big[Y_{l_x}({\widehat {\bf x}_i}) 
Y_{l_y}({\widehat {\bf y}_i}) \big]_\Lambda 
Y_{l_z}({\widehat {\bf z}_i}) 
  \Big]_{J M}, \nonumber \\
(i=1, ..., 12)
\end{eqnarray}
\begin{eqnarray}
 \Phi^{({H})}_{\alpha_{H}}( {\bf x}_i, {\bf y}_i, {\bf z}_i ) 
= \phi^{(^{\rm cos}_{\rm sin})}_{n_x l_x}(x_i) \,
  \psi_{n_y l_y}(y_i) \,
  \varphi_{n_z l_z}(z_i) \nonumber \\
\times \Big[ \big[Y_{l_x}({\widehat {\bf x}_i}) 
Y_{l_y}({\widehat {\bf y}_i}) \big]_\Lambda 
Y_{l_z}({\widehat {\bf z}_i}) 
  \Big]_{J M}, \nonumber \\
(i=13, ..., 18)
\end{eqnarray}
where $\alpha_{K}$ specifies the set 
\begin{eqnarray}
\!\!\!\alpha_{K} \equiv \mbox{
 \{cos or sin}, \omega,
n_x l_x, n_y l_y, n_z l_z , \Lambda,JM \},
\end{eqnarray}
which is the same for the components  $i= 1, ..., 12$;
and similarly for $\alpha_{H}$,
for all $i=13, ..., 18$.
$J$ is the total angular momentum  and
$M$ is its $z$ component. 
In this paper, we consider the tetramer bound states with $J=0$.
Therefore,
the totally symmetric four-body wave function
requires (i)~$l_x={\rm even}$,  $l_y+l_z={\rm even}$ and
$\Lambda=l_z$ for the $K$-type basis and 
(ii)~$l_x={\rm even}$,  $l_y={\rm even}$ and
$\Lambda=l_z={\rm even}$ for the $H$-type basis.

In Eqs. (2.8) and (2.9), the radial functions are assumed as
\begin{eqnarray}
&&  \phi^{(^{\rm cos}_{\rm sin})}_{n_x l_x}(x) =
x^{l_x}\:e^{- (x/x_{n_x})^2} \! \times \!
\left\{ \begin{array}{ll}
  \!\!{\rm cos}\, \omega (x/x_{n_x})^2 & \\
  \!\!{\rm sin}\, \omega (x/x_{n_x})^2 & 
  \end{array}
 \!\!\!\! ,\right. 
 \\
&& \psi_{n_y l_y}(y)=
y^{l_y}\:e^{- (y/y_{n_y})^2}, \: \;  \\ 
&& \varphi_{n_z l_z}(z)=
z^{l_z}\:e^{-(z/z_{n_z})^2} \: \, \;   
\end{eqnarray}
with  geometric sequences of the Gaussian ranges:
\begin{eqnarray}
&& x_{n_x}= x_1\, a_x^{n_x-1}
\quad \:(n_x=1, ..., n_x^{\rm max})\:,  \\
&& y_{n_y}=y_1\, a_y^{n_y-1} \,
\quad \:(n_y=1, ..., n_y^{\rm max})\:,\; \\ 
&& z_{n_z}=z_1\, a_z^{n_z-1} \:
\quad \:(n_z=1, ..., n_z^{\rm max})\:.\;  
\end{eqnarray}
%

It shoud be emphasized that  the GEM few-body calculations 
need neither the introduction of 
any {\it a priori} pair correlation function 
(such as the Jastrow function) 
nor separation of the coordinate space into \mbox{$x<r_c$} 
and \mbox{$x>r_c$}, 
with $r_c$ being the radius of a strongly repulsive core potential.  
Proper short-range correlation and  asymptotic
behavior of the total wave function 
are  {\it  automatically} obtained
by solving the Schr\"{o}dinger equation (2.1)  
using the above  basis functions for {\it ab initio} calculations.

\vskip 0.1cm

We take the same three- and four-body Gaussian basis functions 
as those employed in paper I. The numbers of the 
total bases are 4400 for the trimer and 29056 for the tetramer; 
the bases 
range from very compact to very diffuse with the Gaussian ranges 
in geometric sequences.

\section{The PCKLJS potential and $^4$H\lowercase{e} trimer and tetramer}

The currently most accurate {\it ab initio} potential,
the PCKLJS~\cite{PCKLJS} potential,
is given as a function of the $^4$He pair
separation distance $r$ by
\begin{equation}
V(r) = V_{\rm BO}(r) + V_{\rm ad}(r) + V_{\rm rel}(r) + V_{\rm QED}(r),
\end{equation} 
which are composed of the nonrelativistic BO 
potential ($V_{\rm BO})$ and the leading order coupling of the 
electronic and nuclear motions, that is,
the adiabatic correction ($V_{\rm ad}$), 
relativistic corrections ($V_{\rm rel}$),
and quantum electrodynamics corrections ($V_{\rm QED}$).
Besides them the Casimir-Polder retardation effect~\cite{Casimir},
denoted as $V_{\rm ret}(r)$, can be optionally added to $V(r)$.
By the PCKLJS potential we mean 
the full $V(r)$ plus the residual retardation 
correction $V_{\rm ret}(r)$. Contributions of the individual corrections
are discussed in Sec.IIIA.

Use of PCKLJS for the dimer~\cite{PCKLJS} 
gives the binding energy $B_2= 1.62 \pm 0.03 $ mK, 
the average separation $\langle r \rangle= 47.1 \pm 0.5\,$ \AA $\,$ and 
the $s$-wave scattering length $a=90.42 \pm 0.92 \,$\AA.
Experimental values of the quantities were
reported~\cite{Grisenti2000} as 
$B_2= 1.1^{+0.3}_{-0.2}$ mK,  
$\langle r \rangle = 52 \pm 4\,$ \AA $\,$ and
$a=104^{+8}_{-18}\,$\AA, but the $B_2$ and 
$a$ were calculated~\cite{Grisenti2000}
from the observed value of  $\langle r \rangle$ 
using rather crude models: $B_2=\hbar^2/(4 m \langle r \rangle^2)$
and $a=2 \langle r \rangle$, where $m$ is  mass of
$^4$He atom. Much better estimates of what should be the values
of $B_2$ and $a$ corresponding to the experimental
$\langle r \rangle$ were recently obtained in Ref.~\cite{Cencek2011},
a follow-up paper to Ref.~\cite{PCKLJS}, to be 
$B_2= 1.3^{+0.25}_{-0.19}$ mK  
and $a=100.2^{+8.0}_{-7.9}\,$\AA, which are substantially close
to and nearly consistent with the \mbox{{\it ab initio}} 
results~\cite{PCKLJS} mentioned above. 

Calculated binding energies of the trimer and tetramer
for the PCKLJS potential are
$B_3^{(0)}= 131.84$ mK, $B_3^{(1)}= 2.6502$ mK (1.03 mK below the
dimer), and $B_4^{(0)}= 573.90$ mK,
$B_4^{(1)}= 132.70$ mK (0.86 mK below the trimer ground state).
Some of the mean values of the trimer (tetramer) 
ground and excited states,
as well as the binding energies mentioned above 
are summarized in Table~\ref{table:PCKLJS-trimer} 
(Table~\ref{table:PCKLJS-tetramer}).
The PCKLJS potential gives slightly deeper binding of
the trimer and tetramer than the LM2M2 potential does~\cite{Hiyama2012}.


\subsection{Spatial structure of the tetramer}

We discuss the spatial structure of the tetramer excited state.
In the study of four-boson states and their connection to
the Efimov physics,
the authors of Refs.~\cite{Hammer07,Stecher09} 
predicted that below each Efimov trimer
a pair of tetramer states ($J^\pi=0^+$) should exist
and that the shallower member of the lowest-lying pair
is dominantly composed of
the ground-state trimer and a distant atom.
It  is shown, in the calculations by Lazauskas and 
Carbonell~\cite{Carbonell06} 
and by the present authors~\cite{Hiyama2012} using the 
realistic LM2M2 potential, that
the above prediction is realized in the two bound states 
of the $^4$He tetramer below the trimer ground state.

\begin{figure}[h]
\begin{center}
\epsfig{file=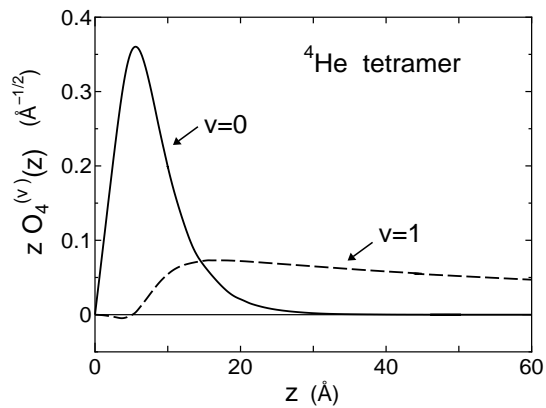,width=7.2cm,height=5.4cm}
\end{center}
\caption{Overlap function ${\cal O}_4^{(v)}(z)$ in Eq.~(3.2) 
between the tetramer state ($v=0, 1$) and
the trimer ground state
as a function of the atom-trimer distance $z$.
}
\label{fig:tetra-red-short}
\end{figure}

The structure of the $^4$He tetramer excited state
is seen essentially in Fig.~\ref{fig:tetra-red-short} for the
overlap function ${\cal O}_4^{(v)}(z)$
between  a tetramer state $\Psi_4^{(v)} (v=0,1)$ 
and the  trimer ground state $\Psi_3^{(0)}$ which
is defined as  a function of the distance $z$ between
the trimer and the fourth atom:
\begin{equation}
{\cal O}_4^{(v)}(z_1)\,Y_{00}({\widehat {\bf z}}_1) =  
\langle \,  \Psi_3^{(0)} \, | \,
\Psi_4^{(v)} \,\rangle_{{\bf x}_1, {\bf y}_1} .
\end{equation}
Figure~\ref{fig:tetra-red-short} indicates that the fourth atom is located 
in the trimer core region in the tetramer 
ground state but is far from the trimer in the
excited state.  This is also understood from the fact that,
in Tables~\ref{table:PCKLJS-trimer} and ~\ref{table:PCKLJS-tetramer}, 
the binding energy and the quantities 
$\langle T \rangle$ and $\langle V \rangle$
for the tetramer excited state
are very close to those for the trimer ground state.

%
\begin{table} [t]  
\begin{center}
\caption{The binding energies $B_3^{(v)} (v=0,1)$ and mean values 
\mbox{of the $^4$He trimer ground} 
and excited states using the PCKLJS 
potential~\cite{PCKLJS} including all the corrections.
$r_{ij}$ stands for interparticle
distance and $r_{i{\rm G}}$ is the distance of a particle from the
center of mass of the trimer. 
$C_3^{(v)}$ is the asymptotic normalization coefficient
defined by Eq.(2.25) in paper~I.
The conversion constant $\frac{\hbar^2}{m}=12.11928$ K\AA$^2$ is taken.
}
\begin{tabular}{ccrrr} 
\hline 
\hline
\noalign{\vskip 0.1 true cm} 
Trimer &  & Ground state 
 & & Excited state  \\
\noalign{\vskip -0.1 true cm} 
(PCKLJS)&  & $(v=0)\;\;\;\;\;$ & & $(v=1)\;\;\;\;\;$  \\
\noalign{\vskip 0.1 true cm} 
\hline 
\noalign{\vskip 0.1 true cm} 
  $B_3^{(v)}$  (mK) & $\quad$
&   131.84 $\quad$    & $\quad$ &  2.6502  $\quad$   \\
 $\langle T \rangle $  (mK)   & $\quad$
&  1694.0 $\quad$    & &  132.0   $\quad$      \\
 $\langle V \rangle $  (mK) & $\quad$
&  $-1825.8$ $\quad$   &    & $-134.7$    $\quad$         \\
 $ \sqrt{ \langle r^2_{ij} \rangle  }$ (\AA) & $\quad$
&  $ 10.83$   $\quad$    & & 100.4  $\quad$  \\
 $  \langle r_{ij} \rangle  $ (\AA) & $\quad$
&  $9.53$  $\quad$      &  & 81.15   $\quad$   \\
 $  \langle r_{ij}^{-1} \rangle  $ (\AA$^{-1}$) & $\quad$
&  $ 0.136 $ $\quad$           &  &  0.0276 $\quad$    \\
 $  \langle r_{ij}^{-2} \rangle  $ (\AA$^{-2}$) & $\quad$
&  $ 0.0231 $ $\quad$            &  & 0.00231   $\quad$    \\
 $ \sqrt{ \langle r_{i{\rm G}}^{2} \rangle } $ (\AA) & $\quad$
&  6.254  $\quad$       & &  57.95   $\quad$     \\
  $C_3^{(v)}$(\AA$^{-\frac{1}{2}})$ & $\quad$ 
&  0.592   $\quad$      & &  0.178   $\quad$      \\
\noalign{\vskip 0.1 true cm} 
\hline
\label{table:PCKLJS-trimer}
\end{tabular}
\end{center}
\end{table}

%
\begin{table} [t]  
\begin{center}
\caption{The binding energies $B_4^{(v)} (v=0,1)$ and mean 
values of the $^4$He tetramer ground 
and excited states using the PCKLJS
potential~\cite{PCKLJS} including all the corrections.
$r_{ij}$ stands for interparticle
distance and $r_{i{\rm G}}$ is the distance of a particle from the
center-of-mass of the tetramer. 
$C_4^{(v)}$ is the asymptotic normalization coefficient
defined by Eq.(3.22)  in paper~I.
The conversion constant $\frac{\hbar^2}{m}=12.11928$ K\AA$^2$ is taken.
}
\begin{tabular}{ccrrr} 
\hline 
\hline
\noalign{\vskip 0.1 true cm} 
Tetramer &  & Ground state 
 & & Excited state  \\
\noalign{\vskip -0.1 true cm} 
(PCKLJS)&  & $(v=0)\;\;\;\;\;$ & & $(v=1)\;\;\;\;\;$  \\
\noalign{\vskip 0.1 true cm} 
\hline 
\noalign{\vskip 0.1 true cm} 
  $B_4^{(v)}$  (mK) & $\quad$
&   573.90 $\quad$    & $\quad$ &  132.70  $\quad$   \\
 $\langle T \rangle $  (mK)   & $\quad$
&  4340.4      $\quad$    & &  1673.4       $\quad$      \\
 $\langle V \rangle $  (mK) & $\quad$
&  $ -4914.3      $ $\quad$   &    & $ -1806.1 $    $\quad$  \\
 $ \sqrt{ \langle r^2_{ij} \rangle  }$ (\AA) & $\quad$
&  $  8.35  $   $\quad$    & & 54.5    $\quad$  \\
 $  \langle r_{ij} \rangle  $ (\AA) & $\quad$
&  $ 7.65  $  $\quad$      &  & 35.8      $\quad$   \\
 $  \langle r_{ij}^{-1} \rangle  $ (\AA$^{-1}$) & $\quad$
&  $  0.156    $ $\quad$           &  &   0.0797     $\quad$    \\
 $  \langle r_{ij}^{-2} \rangle  $ (\AA$^{-2}$) & $\quad$
&  $ 0.0288    $ $\quad$            &  &  0.0119      $\quad$    \\
 $ \sqrt{ \langle r_{i{\rm G}}^{2} \rangle } $ (\AA) & $\quad$
&   5.12 $\quad$       & &   33.0  $\quad$     \\
  $C_4^{(v)}$(\AA$^{-\frac{1}{2}})$ & $\quad$ 
&   2.1    $\quad$      & &   0.10     $\quad$      \\
\noalign{\vskip 0.1 true cm} 
\hline
\label{table:PCKLJS-tetramer}
\end{tabular}
\end{center}
\end{table}

\subsection{Relativistic and QED corrections}

The first four columns of Table~\ref{table:PCKLJS-individual}
list the calculated dimer binding energy
and the average interparticle distance 
at each level of \mbox{theory [PCKLJS (a) to (h)]}, 
showing the contributions of
$V_{\rm BO}, V_{\rm ad}, V_{\rm rel}, V_{\rm QED}$ and
the retardation corrections (\mbox{denoted} as ``{\rm r.c.}''), which
are different at different 
levels of theory~\footnote{For example,  
the retardation correction for $V_{\rm BO}$ is 0.16 mK, but that for 
$V_{\rm BO}+V_{\rm ad}+V_{\rm rel}+V_{\rm QED}$ is 0.005 mK
in the dimer; namely, the residual contribution
becomes much smaller in the latter. 
See Ref.~\cite{PCKLJS} for the details.}.  
The numbers in the first and second columns, 
given by Ref.~\cite{PCKLJS},  are precisely 
reproduced by our calculation.

Using the potentials PCKLJS (a) to (h),
we calculated the binding energies of the ground and excited 
states of the $^4$He trimer and tetramer, 
$B_3^{(0)}, B_3^{(1)}, B_4^{(0)}$ and $B_4^{(1)}$, which are
listed in Table III.
In the tetramer ground-(excited-)state energy, $-B_4^{(0)} (-B_4^{(1)})$, 
the contribution from each correction is as follows:
The retardation correction 
is $+6.87\,(+2.55)$ mK repulsively for $V_{\rm BO}$, but
the residual (remaining)  correction is only $+0.16\,(+0.07)$ mK
for $V_{\rm BO}+V_{\rm ad}+V_{\rm rel}+V_{\rm QED}$. 
Comparing (a) and (c), we see that 
the nonadiabatic correction ($V_{\rm ad}$) is 
$-4.13\, (-1.52)$ mK attractively.
From (c) and (e), the relativistic correction $(V_{\rm rel})$ is known as
$+9.37 \, (+3.48)$ mK.  The QED correction ($V_{\rm QED})$
is $-1.20 \, (-0.46)$ mK from (e) and (g);
the entire correction amounts to +4.20 (+1.57) mK.

We remark that each correction for the tetramer excited-state energy
($-B_4^{(1)}$) 
is approximately the same as the corresponding 
correction for the trimer ground-state energy ($-B_3^{(0)})$.
For example, 
the difference between e) and g), the QED correction, 
is $-0.46\, (-0.46)$ mK
and that between a) and h), the full correction, is $+1.57\, (+1.59)$ mK
for $-B_4^{(1)} (-B_3^{(0)})$.
This is quite reasonable since the 
same explanation in the paragraph below Eq.~(3.2) is applicable.
A similar tendency is seen in the comparison between
the correction for the trimer excited-state energy $(-B_3^{(1)}$) and 
that for the dimer $(-B_2)$; for example,
the QED correction is $-0.036\, (-0.030)$ mK
and the full correction is $+0.12\, (+0.10)$ mK
for $-B_3^{(1)} (-B_2)$.

\begin{table*}[t]
\caption{Calculated binding energies of the $^4$He dimer, 
trimer and tetramer 
using the  PCKLJS potential~\cite{PCKLJS} to demonstrate the contributions of
$V_{\rm BO}, V_{\rm ad}, V_{\rm rel}, V_{\rm QED}$ and
the retardation correction, denoted as "{\rm r.c.}", appropriate for a
given level of theory. PCKLJS-h is the full PCKLJS potential.
$B_2$ and $\langle r \rangle$ are 
the binding energy and the average separation of the dimer, respectively.
Calculated results for the dimer by Ref.~\cite{PCKLJS} are
shown in the first and second columns.
The conversion constant
$\frac{\hbar^2}{m}=12.11928$ K\AA$^2$ is taken.
}
\begin{center}
\begin{tabular}{lcccccccccccccccc} 
\hline \hline
\noalign{\vskip 0.1 true cm} 
    &   & \multicolumn{3}{c} { Dimer~\cite{PCKLJS}}  &    &
              \multicolumn{3}{c}  { Dimer }   
   &   & \multicolumn{3}{c} { Trimer}  &    &
              \multicolumn{3}{c}  { Tetramer }   \\
\noalign{\vskip -0.2 true cm} 
   &    &  \multispan3 {\hrulefill} & \qquad \qquad  &
             \multispan3 {\hrulefill} 
  &\qquad  \qquad   &  \multispan3 {\hrulefill} & \qquad \qquad  &
             \multispan3 {\hrulefill} \\
\noalign{\vskip 0.1 true cm} 
PCKLJS potential~\cite{PCKLJS}   &   & $B_2$ & \qquad  & 
               $\langle r \rangle$  &  \qquad & 
              $B_2$ & \qquad & $\langle r \rangle$   
   & $\qquad$ &       $B_3^{(0)}$ & \qquad & $B_3^{(1)}$  & \qquad & 
             $B_4^{(0)}$ & $\quad$  & $B_4^{(1)}$  \\
\noalign{\vskip 0.03 true cm} 
            &    & (mK)& \qquad  &  (\AA)  &   & 
            (mK) &  \qquad & (\AA)  
           &    & (mK)& \qquad  &  (mK)  &   &
               (mK) & \quad & (mK)  \\
\noalign{\vskip 0.05true cm} 
\hline 
\noalign{\vskip 0.2 true cm} 
(a) $V_{\rm BO}$  & \quad & 
        1.718   & \quad  &  45.77  & \quad & 
        1.7181    & \quad  &  45.77  & \quad & 
        133.43 & \quad  &  2.7724  & \quad & 
        578.10  & \quad &  134.27     \\
(b) $V_{\rm BO} +{\rm r.c.}$  & \quad & 
        1.555  & \quad  &  47.92   & \quad & 
        1.5549   & \quad  &  47.92   & \quad & 
        130.85  & \quad  &  2.5776  & \quad & 
        571.23  & \quad & 131.72     \\
(c) $V_{\rm BO}+V_{\rm ad}$  & \quad & 
        1.816  & \quad  &  44.62   & \quad & 
        1.8160   & \quad  &  44.62   & \quad & 
        134.96  & \quad  &  2.8881  & \quad & 
        582.23       & \quad & 135.79          \\
(d) $V_{\rm BO}+V_{\rm ad}+{\rm r.c.}$  & \quad & 
        1.648  & \quad  &  46.65   & \quad & 
        1.6482   & \quad  &  46.65   & \quad & 
        132.37  & \quad  &  2.6894  & \quad & 
        575.33  & \quad &  133.23   \\
(e) $V_{\rm BO}+V_{\rm ad}+V_{\rm rel}$
   & \quad & 
        1.590  & \quad  &  47.43   & \quad & 
        1.5896   & \quad  &  47.43   & \quad & 
        131.44  & \quad  &  2.6194  & \quad & 
        572.86       & \quad &  132.31          \\
(f) $V_{\rm BO}+V_{\rm ad}+V_{\rm rel}+ {\rm r.c.}$  & \quad & 
        1.611  & \quad  &  47.15   & \quad & 
        1.6105   & \quad  &  47.15   & \quad & 
        131.76  & \quad  &  2.6444  & \quad & 
        573.69        & \quad & 132.62          \\
(g) $V_{\rm BO}+V_{\rm ad}+V_{\rm rel}+V_{\rm QED}$  & \quad & 
        1.620  & \quad  &  47.02   & \quad & 
        1.6200   & \quad  &  47.02   & \quad & 
        131.90  & \quad  &  2.6559  & \quad & 
         574.06  & \quad &  132.77        \\
(h) $V_{\rm BO}+V_{\rm ad}+V_{\rm rel}+V_{\rm QED}+{\rm r.c.}$  & \quad & 
         1.615  & \quad  &  47.09   & \quad & 
        1.6154   & \quad  &  47.09   & \quad & 
        131.84  & \quad  &  2.6502  & \quad & 
        573.90  & \quad &  132.70         \\
\noalign{\vskip 0.1 true cm} 
\hline
\hline
\end{tabular}
\label{table:PCKLJS-individual}
\end{center}
\end{table*}

\section{Universality in $^4$H\lowercase{e} trimer and tetramer}

Universal correlations between observables have been 
studied systematically in four-boson systems interacting
with large scattering length~\cite{Platter04,Hammer07,Stecher09,
Stecher09PRL,Tomio06,Tomio2011PRL,Tomio2011FBS}.
In this section  we investigate the correlations between
the ground- and excited-state binding energies of 
the $^4$He trimer and tetramer.  We calculate the energies using 
\mbox{various realistic} $^4$He-$^4$He interactions which include
the PCKLJS potential and six other potentials 
 LM2M2, TTY, HFD-B, HFD-B3-FCI1,
SAPT96 and  CCSAPT07 mentioned in Sec.~I;
in the last three, we choose the cases in which
the retardation corrections are included.
\begin{table}[t]
\caption{
The binding energy $B_2$ and the 
average interparticle distance $\langle r \rangle$ of the dimer
calculated using the seven $^4$He potentials.  
The conversion constant
$\frac{\hbar^2}{m}=12.11928$ K\AA$^2$ is taken.
The values reported in the literature are also shown,
but the numbers in the parentheses were obtained by using
$\frac{\hbar^2}{m}=12.12$ K\AA$^2$
and those in the square brackets were given with the use of
the $^4$He nuclear mass for $m$. 
The potential names are arranged in the 
increasing order of $B_2$. 
}
\label{table:1}
\begin{center}
\begin{tabular}{cccccccccc} 
\hline \hline
\noalign{\vskip 0.1 true cm} 
 &        & \multicolumn{3}{c} {This work}  &  $\;\;$
            &  \multicolumn{4}{c}  {Other work}    \\
\noalign{\vskip -0.2 true cm} 
  &   &  \multispan3 {\hrulefill} & 
 & \multispan4 {\hrulefill}  \\
\noalign{\vskip 0.1 true cm} 
 Potential &     & $B_2$  & $\;\;$ & $\langle r \rangle$  & & 
$B_2$  & & $\langle r \rangle$ &  \\
\noalign{\vskip 0.03 true cm} 
         &    & (mK)  &  & (\AA)  & & 
(mK) &  & (\AA) & Ref.  \\
\noalign{\vskip 0.05true cm} 
\hline 
\noalign{\vskip 0.2 true cm} 
LM2M2 &  
       $\;$ &   1.3094 & 
      &   51.87   & $\;\;\;$ & (1.3035) &    & 
      (52.00)  & \cite{Hiyama2012} \\
TTY   &  
       $\;$ &   1.3156 &
      &   51.76  & $\;\;\;$ &  (1.3096) &    &  
     (51.89)  & \cite{review-Efimov}  \\
HFD-B3-FCI1 &  
        $\;$ &   1.4475 &
      &   49.52   & $\;\;\;$ &  1.448   &    & 
       49.52  & \cite{Spirko}  \\
CCSAPT07 &  
        $\;$ &   1.5643 &
      &   47.78   & $\;\;\;$ &   1.56   &    &  
      47.8  &  \cite{CCSAPT07}   \\
PCKLJS &  
        $\;$   &   1.6154 &
      &   47.09   & $\;\;\;$ &   1.615  &    &  
      47.09  & \cite{PCKLJS}  \\
HFD-B      &  
        $\;$ &   1.6921 &
      &   46.07  & $\;\;\;$ &    (1.6854)  & &  
(46.18)  & \cite{review-Efimov}\\
SAPT96 &  
        $\;$ &   1.7443 &
      &   45.45  & $\;\;\;$ &  [1.713]&   &  
[45.8]  & \cite{SAPT96b} \\
\noalign{\vskip 0.1 true cm} 
\hline
\hline
\end{tabular}
\label{table:universal-dimer}
\end{center}
\end{table}

\begin{table}[h]
\caption{The binding energies of 
$^4$He trimer and tetramer ground and excited states
calculated with the use of the seven $^4$He potentials.
The conversion constant $\frac{\hbar^2}{m}=12.11928$ K\AA$^2$ is taken.
}
\label{table:1}
\begin{center}
\begin{tabular}{ccccccccc} 
\hline \hline
\noalign{\vskip 0.1 true cm} 
 &        & \multicolumn{3}{c} {Trimer}  &  $\;$
            &  \multicolumn{3}{c}  {Tetramer}   \\
\noalign{\vskip -0.2 true cm} 
  &   &  \multispan3 {\hrulefill} & 
 & \multispan3 {\hrulefill} \\
\noalign{\vskip 0.1 true cm} 
 Potential &     & $B_3^{(0)}$  &  & $B_3^{(1)}$  & & 
$B_4^{(0)}$  & & $B_4^{(1)}$  \\
\noalign{\vskip 0.03 true cm} 
         &    & (mK)  &  & (mK)  & & 
(mK) &  & (mK)  \\
\noalign{\vskip 0.05true cm} 
\hline 
\noalign{\vskip 0.2 true cm} 
LM2M2 &  
       $\;\;\;$ &   126.50 &
      &   2.2779  & $\;\;\;$ &  559.22  &    &  127.42 \\
TTY   &  
       $\;\;\;$ &   126.45 &
      &   2.2844  & $\;\;\;$ &  558.70  &    &  127.37 \\
HFD-B3-FCI1 &  
        $\;\;\;$ &   129.00 &
      &   2.4475  & $\;\;\;$ &  566.12  &    &  129.89 \\
CCSAPT07 &  
        $\;\;\;$ &   131.01 &
      &   2.5890  & $\;\;\;$ &   571.67 &    &  131.88 \\
PCKLJS &  
        $\;\;\;$   &   131.84 &
      &   2.6502  & $\;\;\;$ &  573.90  &    &   132.70 \\
HFD-B      &  
        $\;\;\;$ &   133.08 &
      &   2.7420  & $\;\;\;$ &    577.34&    &  133.94 \\
SAPT96 &  
        $\;\;\;$ &   134.02 &
      &   2.8045 & $\;\;\;$ &  580.01   &    &  134.86   \\
\noalign{\vskip 0.1 true cm} 
\hline
\hline
\end{tabular}
\label{table:resonance}
\end{center}
\end{table}

We first calculated, using the seven potentials,
the binding energy $B_2$ and the average interparticle distance 
$\langle r \rangle$ of the dimer and listed them in
Table~\ref{table:universal-dimer} together with the values reported
in the literature.
The scattering lengths are not listed,  but they range
between 87.92 \AA~\cite{SAPT2} for SAPT96 and 
100.23 \AA~\cite{review-Efimov}
for LM2M2 (90.42 \AA~\cite{PCKLJS} for PCKLJS). 
Here, the names of the seven  potentials
are arranged from the top to the bottom
in the increasing order of $B_2$.

\begin{figure*}[t]
\begin{center}
\begin{minipage}[t]{8 cm}
\epsfig{file=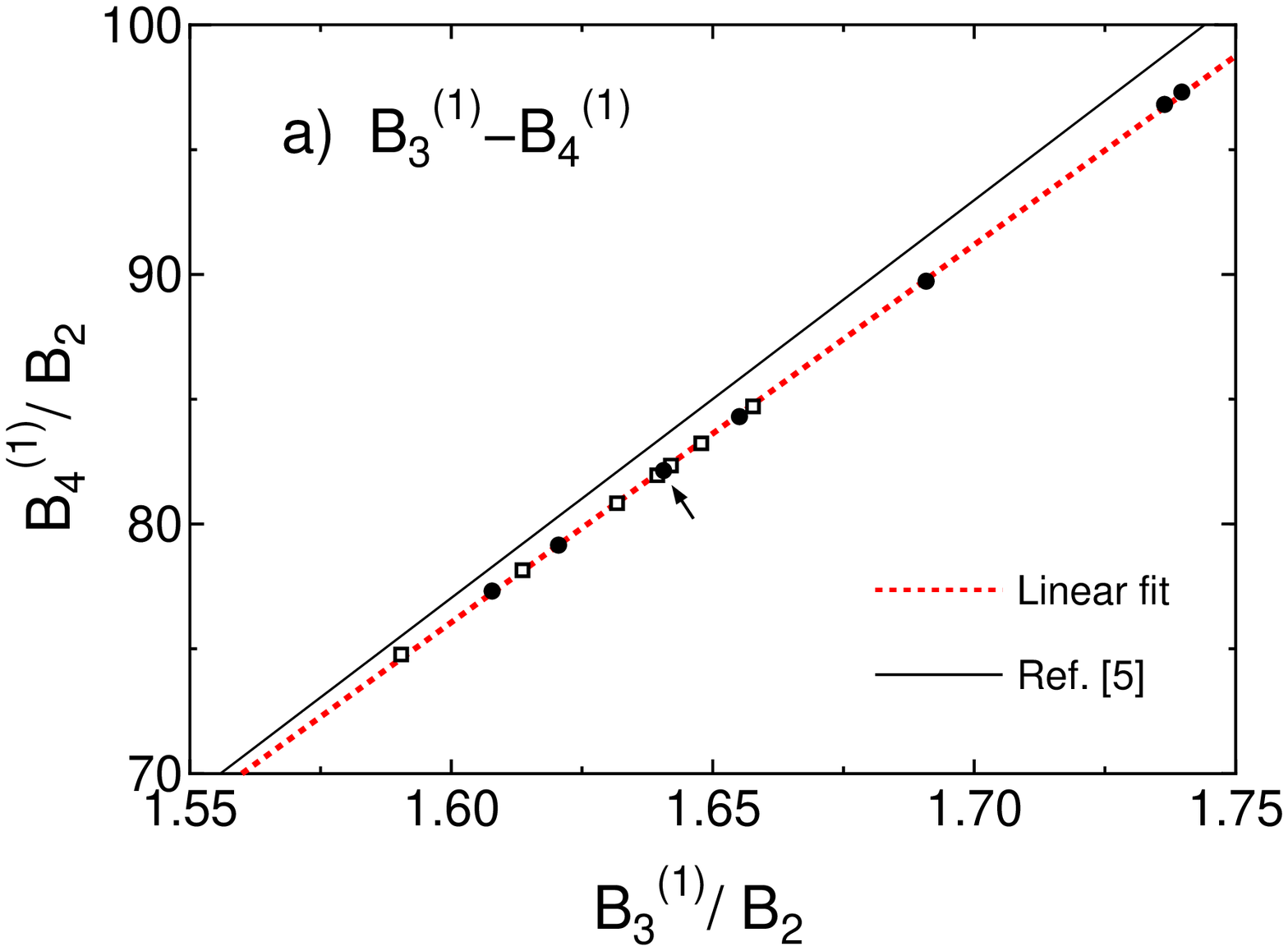,scale=0.43}
\end{minipage}
\hspace{\fill}
\begin{minipage}[t]{9 cm}
\epsfig{file=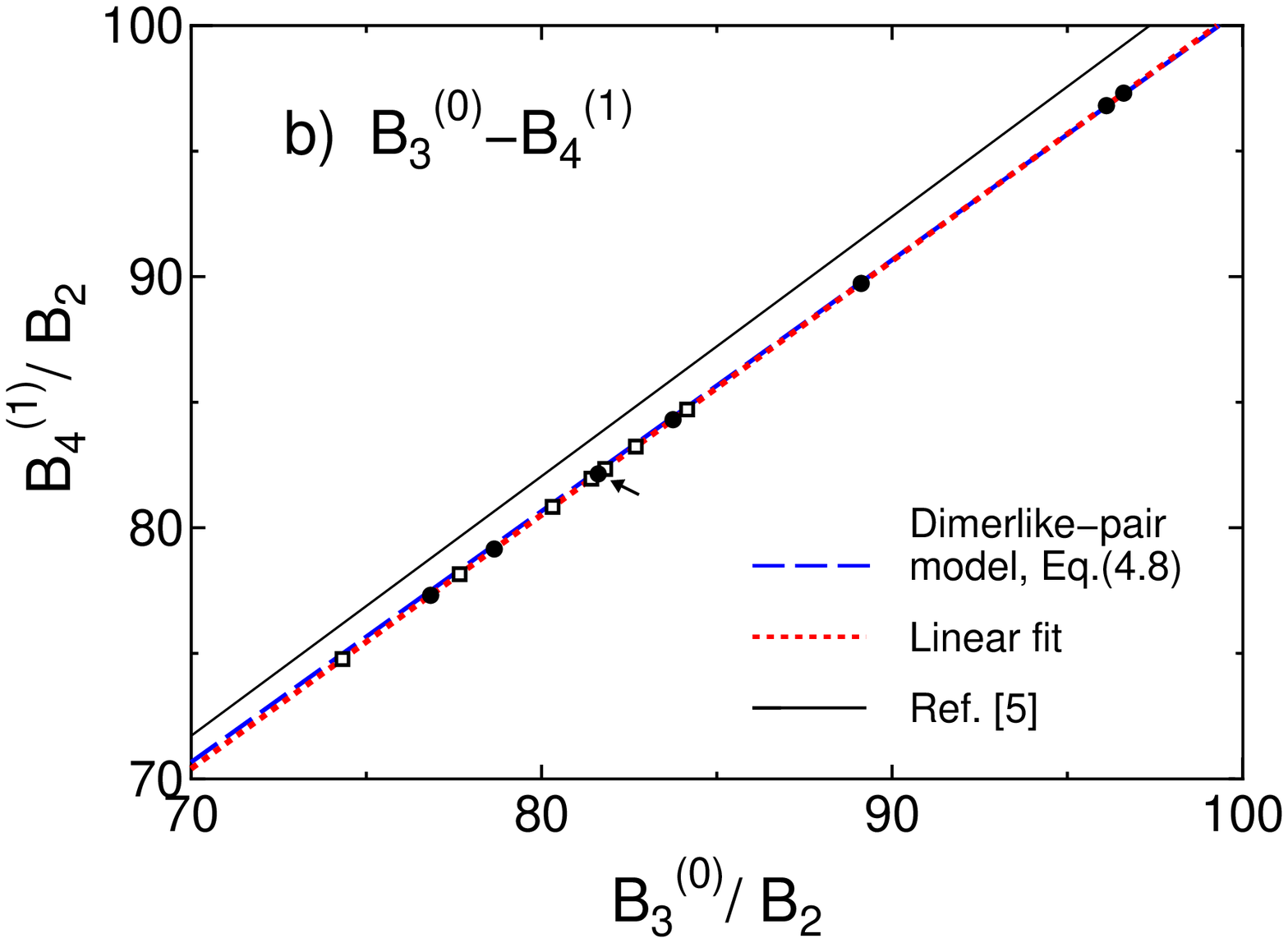,scale=0.43}
\end{minipage}
\begin{minipage}[t]{8 cm}
\epsfig{file=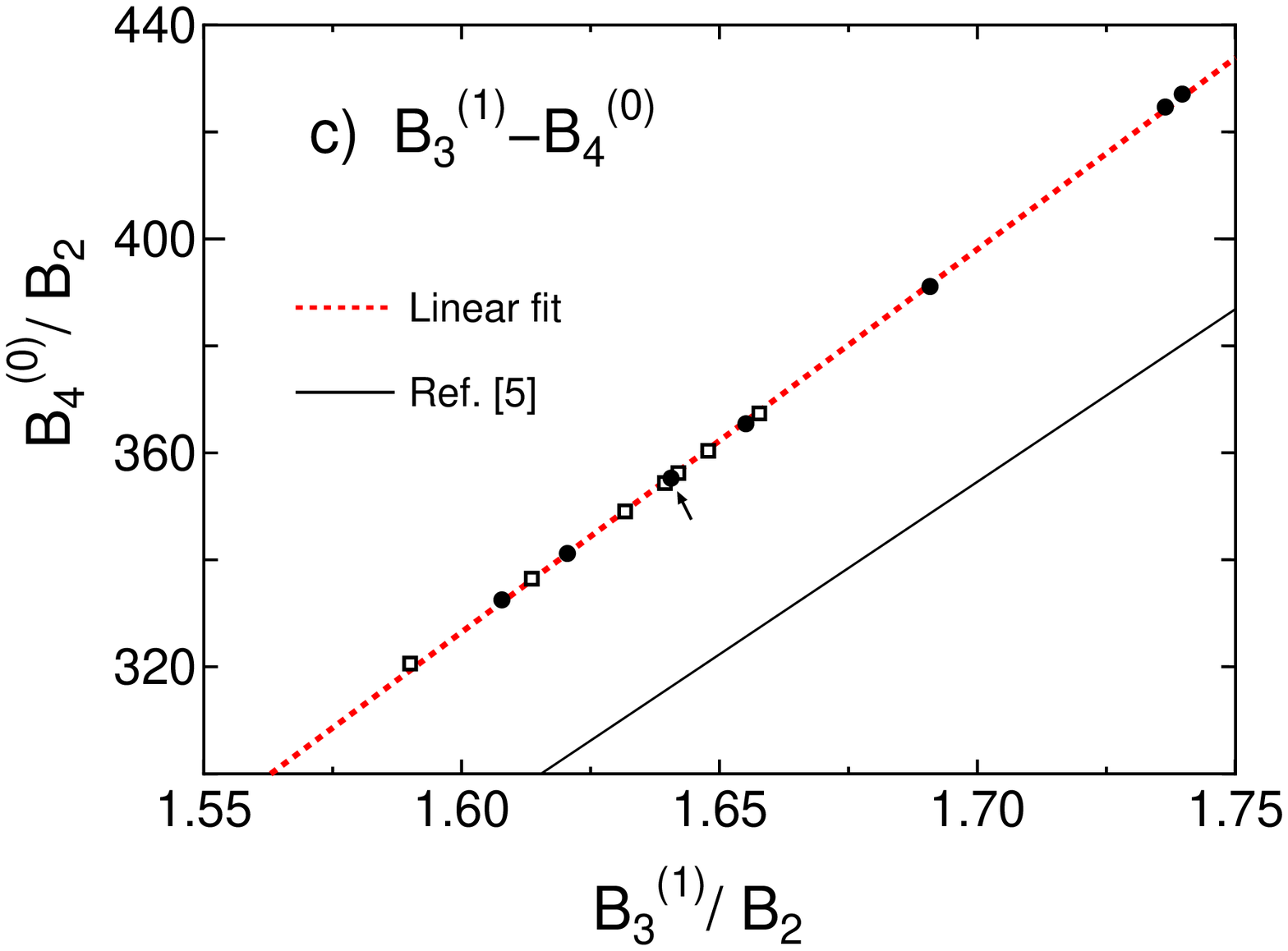,scale=0.43}
\end{minipage}
\hspace{\fill}
\begin{minipage}[t]{9 cm}
\epsfig{file=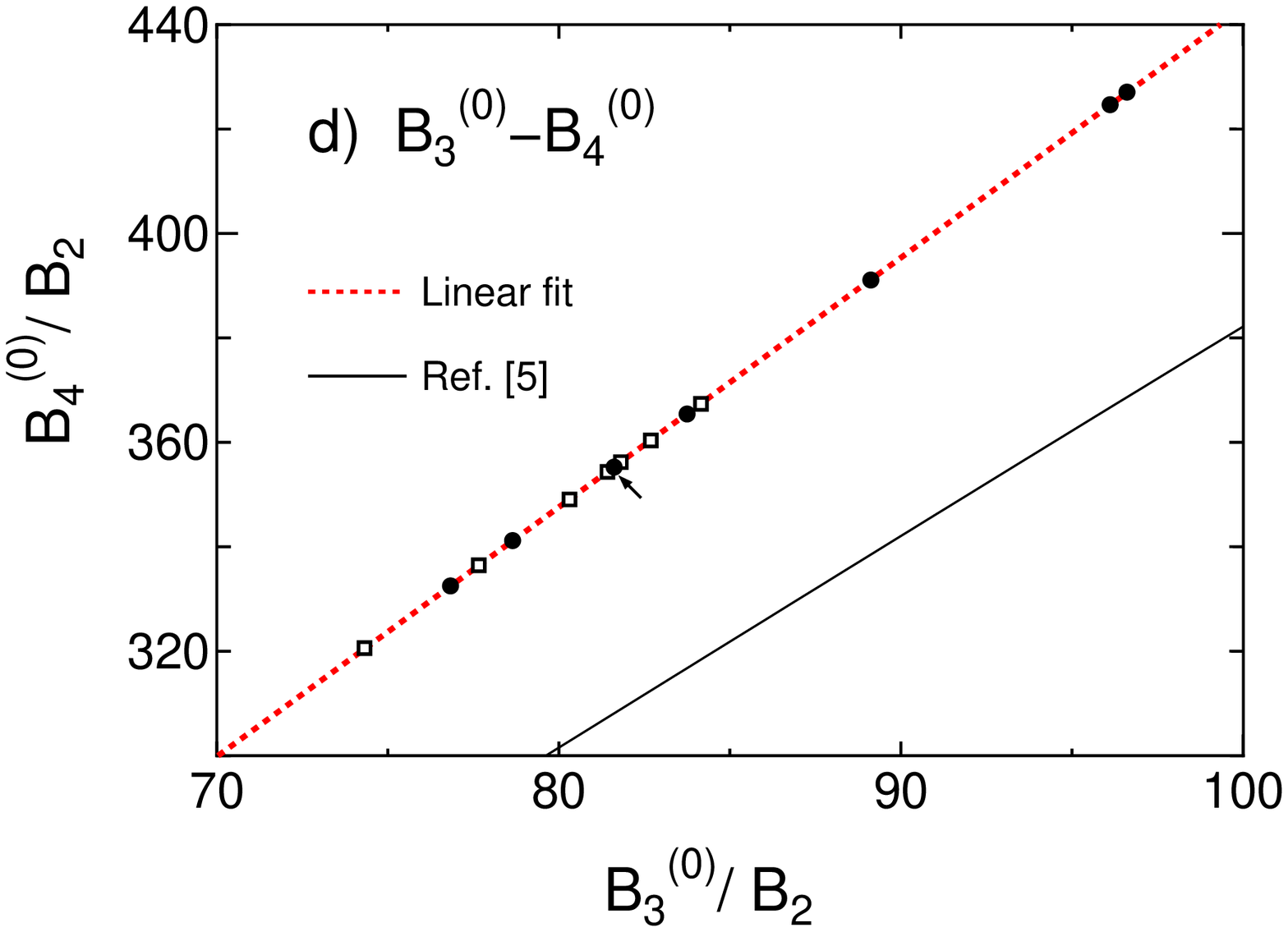,scale=0.43}
\end{minipage}
\end{center}
\caption{(Color online).
The correlations between the ground- and excited-state
binding energies of the $^4$He trimer and tetramer.
(a) $B_3^{(1)}$-$B_4^{(1)}$, (b) $B_3^{(0)}$-$B_4^{(1)}$,
(c) $B_3^{(1)}$-$B_4^{(0)}$ and (d) $B_3^{(0)}$-$B_4^{(0)}$
correlations.
The energies are normalized by the dimer energy $B_2$.
All the 14 data points are obtained by  the present calculation 
for various $^4$He potentials. The seven closed circles,
from the right to the left, denote the results
for the seven potentials in Table V from LM2M2 down to SAPT96,
respectively; 
the one designated by an arrow is for PCKLJS.
The seven open squares, from the left to the right,
show the calculation for each level of the PCKLJS potential in
Table I in the order of PCKLJS (c), (a), (d), (g), (f), (e), and 
(b), respectively.
The dotted (red) linear line is
the linear least squares fit to the 14 data points;
see Eqs.~(4.1)--(4.4). 
The dashed (blue) line  
in panel (b) is the prediction by the dimerlike-pair model, Eq.~(4.8).
The solid line, taken from Fig.~4 and Eqs.~(39)--(42) 
in Ref.~\cite{Platter04},
shows the universal scaling curve obtained by
the leading-order effective theory for the $^4$He trimer and tetramer.
}
\label{fig:scal-34}
\end{figure*}

The $^4$He trimer and tetramer 
ground- and excited-state binding energies,
$B_3^{(0)}, B_3^{(1)}, B_4^{(0)}$ and $B_4^{(1)}$,  
are calculated with those potentials and are listed in Table V.
The values of each binding energy appear 
in the increasing order as $B_2$ does in Table IV except for
$B_3^{(0)}, B_4^{(0)}$ and $B_4^{(1)}$ for LM2M2 and TTY.
This exception is reasonable   
because 
TTY is slightly more attractive for \mbox{$r>2.65$ \AA}$\,$than LM2M2,
but slightly more repulsive for \mbox{$r<2.65$ \AA}; namely,
it is possible that TTY generates larger binding energies than LM2M2
in loosely bound systems (the dimer and the trimer excited states)
but brings about smaller binding energies in 
compactly bound 
systems (the trimer and tetramer ground states and
the tetramer excited state that is dominantly \mbox{composed} of
the compact trimer ground state and a distant $^4$He atom).
We note that if all the binding energies are normalize by $B_2$,
they  appear in the increasing order in Table V.

\subsection{Linear correlations}

\begin{figure*}[t]
\begin{center}
\begin{minipage}[t]{8 cm}
\epsfig{file=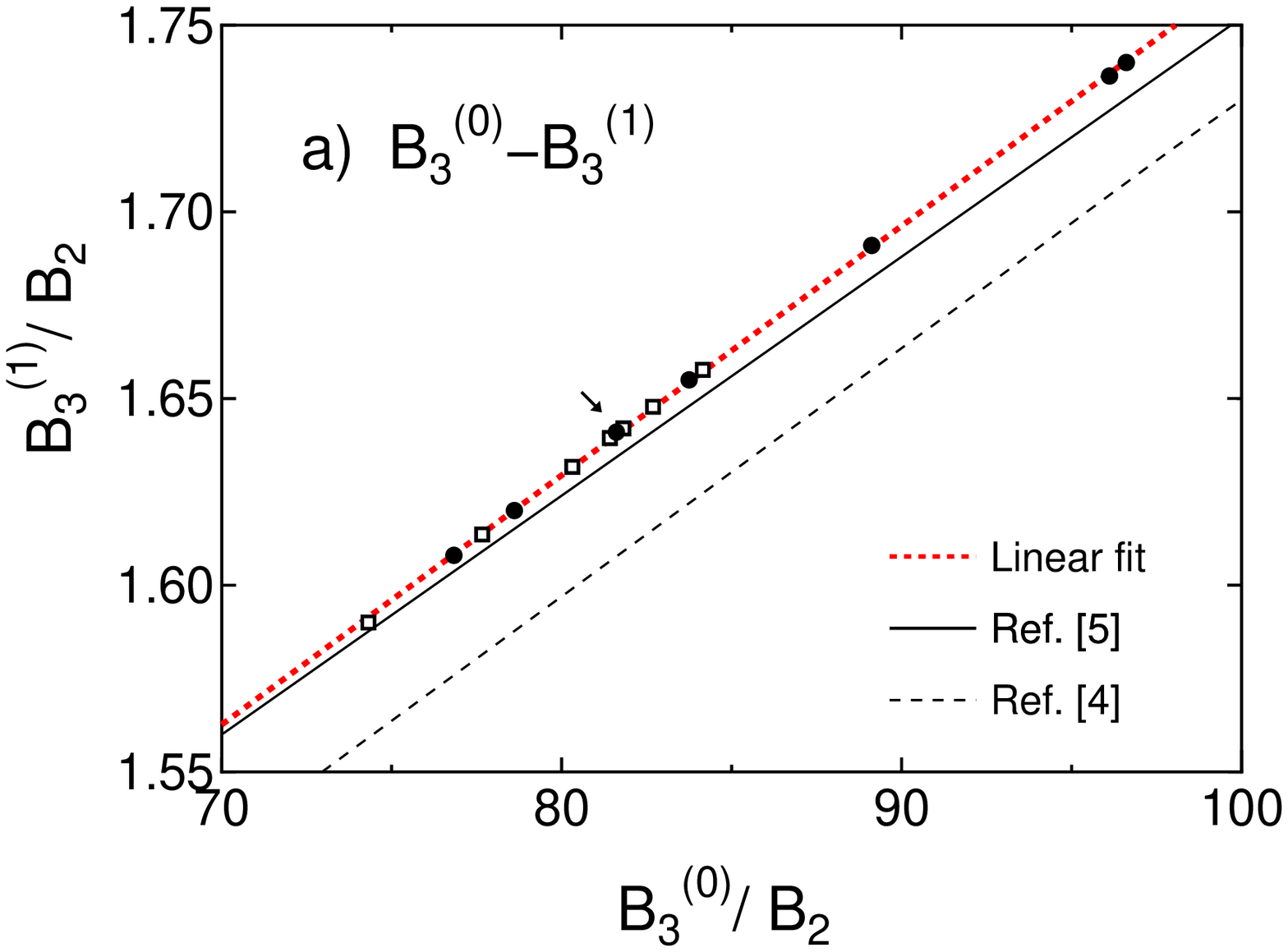,scale=0.44}
\end{minipage}
\hspace{\fill}
\begin{minipage}[t]{9 cm}
\epsfig{file=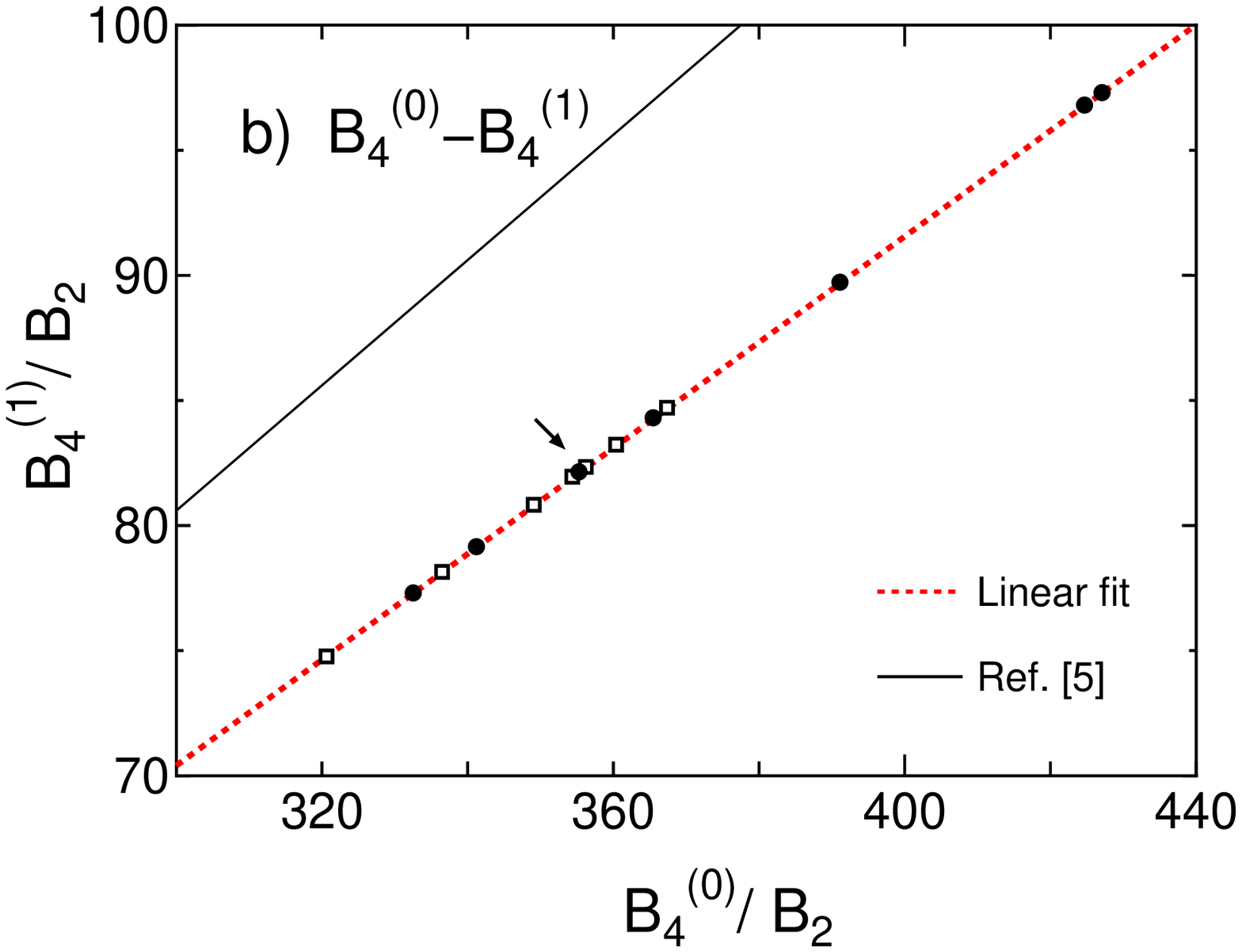,scale=0.44}
\end{minipage}
\end{center}
\caption{(Color online). 
(a) $B_3^{(0)}$-$B_3^{(1)}$ and  (b) $B_4^{(0)}$-$B_4^{(1)}$
correlations.
Meanings of the data points (the present calculation) are 
the same as in Fig.~\ref{fig:scal-34}.
The dotted (red) linear line is the linear least squares fit 
to the 14 data points; see Eqs.~(4.5) and (4.6). 
The solid line shows the universal scaling curve 
obtained by the leading-order effective theory
for the $^4$He atoms; we derived the line 
from Eqs.~(39)--(42) of Ref.~\cite{Platter04} (see footnote [49]). 
The dashed line in (a) is another universal scaling curve obtained in
Ref.~\cite{Braaten03}.
}
\label{fig:scal-3344}
\end{figure*}

Correlations between 
the  binding energies in three- and four-body systems
were first observed in nuclear physics and are known as 
the Tjon line~\cite{Tjon}, which refers to the approximately
linear correlation between the binding energies of 
the triton and the $\alpha$ particle for various nucleon-nucleon potentials.
Recently, the nuclear Tjon line was discussed in the context of 
the effective field theory of short-range interactions and low-momentum 
nucleon-nucleon potentials~\cite{Tjon-N,Tjon-P}.
The Tjon lines for the $^4$He trimer and tetramers
were investigated in Refs.~~\cite{Braaten03,Platter04}
over the range of binding energies relevant to $^4$He atoms
on the basis of the leading-order effective theory.
However, due to the scarce calculation of the
$^4$He tetramer excited-state binding energy $B_4^{(1)}$
with the realistic $^4$He potential at that time, 
the correlations associated with $B_4^{(1)}$ remained unexplored.

We consider all six kinds of the correlations 
between two of the four binding energies,
$B_3^{(0)}, B_3^{(1)}, B_4^{(0)}$ and $B_4^{(1)}$,
that are calculated using 
the seven $^4$He potentials in Table V and the 
seven potentials of PCKLJS (a) to (g) in Table I.
The binding energies are normalized by $B_2$ , which is 
different for different $^4$He potentials; this is due to
the fact that the experimental value of the dimer binding energy
has not been precisely obtained
as mentioned at the beginning of Sec.~III.

Figures~\ref{fig:scal-34}(a) to~\ref{fig:scal-34}(d) 
illustrate the 
\mbox{$B_3^{(1)}$-$B_4^{(1)}$}, \mbox{$B_3^{(0)}$-$B_4^{(1)}$},
\mbox{$B_3^{(1)}$-$B_4^{(0)}$} and \mbox{$B_3^{(0)}$-$B_4^{(0)}$}
correlations, respectively.
The \mbox{14 data} points are given by  the present \mbox{calculation} 
for various \mbox{14 potentials} mentioned above. 
The dotted (red)  lines
in Figs.~\ref{fig:scal-34}(a) to \ref{fig:scal-34}(d) 
are obtained by the linear least squares fitting
to the data points and are represented 
by the following equations, respectively:
\begin{eqnarray}
&&\frac{B_4^{(1)}}{B_2} = 151.4 \frac{B_3^{(1)}}{B_2}-166.1 , \quad
 70 \lesssim \frac{B_3^{(1)}}{B_2} \lesssim 100, \\ \nonumber \\
&& \frac{B_4^{(1)}}{B_2} =  1.011 \frac{B_3^{(0)}}{B_2}-0.3694 , \;\;
 1.5 \lesssim \frac{B_3^{(0)}}{B_2} \lesssim 1.8, \qquad \\ \nonumber\\
&&\frac{B_4^{(0)}}{B_2} =  715.9 \frac{B_3^{(1)}}{B_2}-819.0 , \quad
 70 \lesssim \frac{B_3^{(1)}}{B_2} \lesssim 100, \\ \nonumber\\
&& \frac{B_4^{(0)}}{B_2} =   4.778 \frac{B_3^{(0)}}{B_2}-34.64, \quad
 1.5 \lesssim \frac{B_3^{(0)}}{B_2} \lesssim 1.8. \qquad  
\end{eqnarray}

Figures~\ref{fig:scal-3344}(a) and 4(b) plot the 
\mbox{$B_3^{(0)}$-$B_3^{(1)}$} and  \mbox{$B_4^{(0)}$-$B_4^{(1)}$}
correlations.
With the least squares method,
the data points  are fitted 
by the dotted (red) lines  that are represented by
\begin{eqnarray}
&& \!\!\!\!\!\!\!\!\!\!\! \frac{B_3^{(1)}}{B_2} =  0.006679 
\frac{B_3^{(0)}}{B_2}+1.095, \;\;
 70 \lesssim \frac{B_3^{(0)}}{B_2} \lesssim 100 ,\quad \\ \nonumber\\
&&\!\!\!\!\!\!\!\!\!\! \frac{B_4^{(1)}}{B_2} =  0.2116 
\frac{B_4^{(0)}}{B_2}+6.961,\quad 
 300 \lesssim \frac{B_4^{(0)}}{B_2} \lesssim 440 .
\end{eqnarray}

In Figs.~\ref{fig:scal-34} and \ref{fig:scal-3344}, the scattering of 
the data points about the fitted linear line is very small:
Representing the data points by $\{ (x_i, y_i), i=1,...,14\}$
and the fitted linear function by $y=f(x)$,
we define relative deviation at each $x_i$ by $|y_i - f(x_i)|/y_i$.
The average values of the relative deviation in
Figs.~3(a), 3(b), ...., 4(b) are respectively  
0.093\%, 0.0032\%, 0.11\%, 
0.019\%, 0.030\% and
0.015\%.
We remark that, among Eqs.~(4.1)--(4.6), any three equations 
can be reproduced by the other three (linearly dependent)  
with very small errors.  This comes from the 
fact that  the six kinds of  correlations are 
all linear for various potentials.

It is unexpected that 
all the calculated results (the data points)  fall  so strictly
on a straight line over the range of binding energies 
relevant for $^4$He atoms; 
we emphasize that the results are obtained by using 
{\it different} potentials, not by changing  parameter(s)
in a specific potential.

It is of interest to note that the slope of  
the dotted (red) line in Fig.~3(d) for the correlation between
trimer and tetramer ground-state binding energies is 4.778 [see Eq.~(4.4)],
which is close to the slope of the nuclear Tjon line 
($\approx 5.0$~\cite{Tjon-N}) 
for the correlation between three- and four-nucleon binding energies 
using various nucleon-nucleon potentials.

Another similarity between the $^4$He tetramer and the 
$^4$He nucleus is seen in the comparison of the
overlap function ${\cal O}_4^{(v)}(z) $ in Fig.~2 ($^4$He tetramer)
with that in Fig.~5 ($^4$He nucleus).
The behavior of ${\cal O}_4^{(v)}(z) $ is quite resemble 
each other although the sizes of the systems
are very different.
The first excited $0^+$ state of the $^4$He nucleus is known to be
composed of the three-nucleon core and 
a loosely coupled nucleon~\cite{Hiyama03second}.
The observed cross sections of the electron inelastic scattering, 
which drastically excites the compact ground state 
to the diffuse excited state, 
is well explained by
the GEM four-body calculation by the present authors~\cite{Hiyama03second}.

In the study of weakly bound four-boson states (not specifically for
$^4$He atoms) 
at the unitary limit, 
von Stecher {\it et al.}~\cite{Stecher09} obtained
\mbox{$B_4^{(0)}/B_3^{(0)}\approx4.58$} and 
\mbox{$B_4^{(1)}/B_3^{(0)}\approx1.01$}, while
Deltuva~\cite{Deltuba10} gave
\mbox{$B_4^{(0)}/B_3^{(0)}\approx4.611$} and 
\mbox{$B_4^{(1)}/B_3^{(0)}\approx1.0023$}, and 
Hadizadeh {\it et al.}~\cite{Tomio2011PRL} reported
\mbox{$B_4^{(0)}/B_3^{(0)}\approx4.6$}.
In the $^4$He atoms, we can estimate the ratio, 
approximately from Eqs.~(4.4) and (4.2), as
\mbox{$B_4^{(0)}/B_3^{(0)}=4.778-34.64 B_2/B_3^{(0)}\approx$}~4.3--4.4
and 
$B_4^{(1)}/B_3^{(0)}=1.011-0.3694$ 
\mbox{$B_2/B_3^{(0)}\approx$}~1.006--1.007
over the range of binding energies relevant
for the $^4$He atoms.

\begin{figure}[h]
\epsfig{file=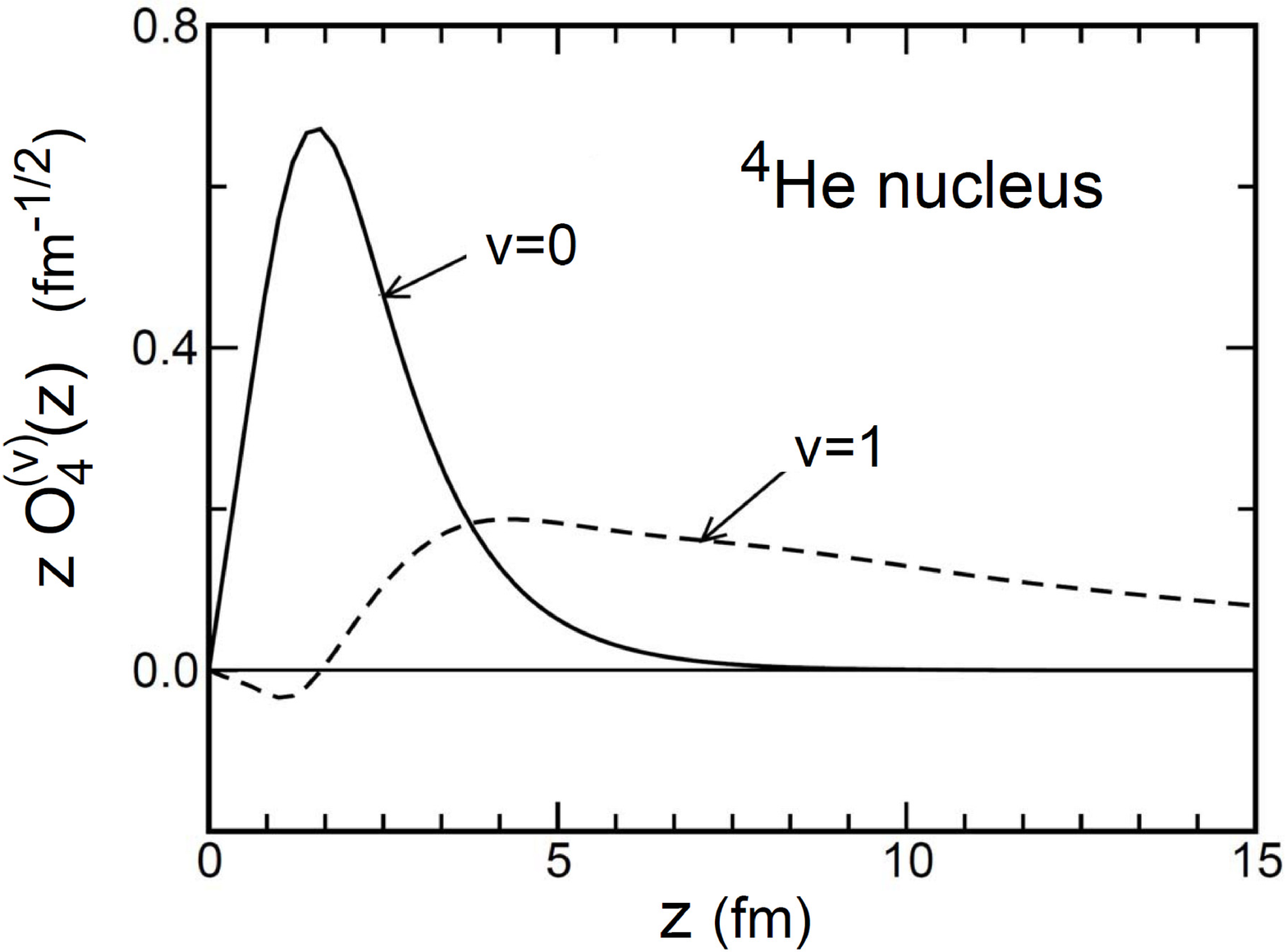,width=7.2 cm,height=5.4 cm}
\caption{
Overlap function ${\cal O}_4^{(v)}(z)$   between
the $^4$He nucleus ($v=0, 1$) and
the three-nucleon ground state
as a function of the  distance $z$
between the three-nucleon core and the fourth nucleon.
Taken from Ref.~\cite{Hiyama03second}.
Note that this figure is quite similar to Fig.~2 for the $^4$He tetramer.
}
\end{figure}

\subsection{Universal scaling functions}

The solid lines in Fig.~\ref{fig:scal-34}
illustrate the universal scaling functions 
relating the tetramer energies to the trimer energies, which were
calculated by the 
leading-order effective theory for the $^4$He atoms; 
the lines are taken from  Eqs.~(39)--(42) and Fig.~4 in 
Ref.~\cite{Platter04}.
To obtain the energies, Platter {\it et al.}~\cite{Platter04} 
constructed an effective $^4$He-$^4$He
potential including both  two- and three-body 
contact interactions.
The two parameters of the effective potential were determined 
to reproduce the 
binding energy of the dimer ground state and the trimer excited state.
They solved the three- and four-body Faddeev-Yakubovsky equations
with the effective potential.
Although the $B_3^{(0)}$-$B_3^{(0)}$ and $B_4^{(0)}$-$B_4^{(1)}$
correlations
are not explicitly given in Ref.~\cite{Platter04}, 
we derived the functions for the 
correlations~\footnote{In Ref.~\cite{Platter04}, 
one can obtain an equation to relate $B_3^{(0)}$ and 
$B_3^{(1)}$ by eliminating $B_4^{(0)}$
from Eqs.~(39) and (40)  
and one more equation from Eqs.~(41) and (42)
by eliminating $B_4^{(1)}$. Since the resultant 
two equations are slightly different from each other in the coefficients,
we averaged them and obtained 
$\frac{B_3^{(1)}}{B_2} =  0.006402\frac{B_3^{(0)}}{B_2}+1.112.$
Similarly, we derived 
$\frac{B_4^{(1)}}{B_2} =  0.2504\frac{B_4^{(0)}}{B_2}+5.480.$
They are plotted in Fig.~\ref{fig:scal-3344} with solid lines} 
using Eqs.~(39)--(42) in Ref.~\cite{Platter04}
and plot them in Fig.~\ref{fig:scal-3344} with the solid lines.
The dashed line in Fig.~\ref{fig:scal-3344}(a) is another
universal scaling curve for the $^4$He trimer
given in Fig.~2 of Ref.~\cite{Braaten03}. 

The  solid lines 
in Figs.~\ref{fig:scal-34}(a), ~\ref{fig:scal-34}(b)
and \ref{fig:scal-3344}(a), associated with 
$B_3^{(0)}, B_3^{(1)}$ and $B_4^{(1)}$, 
are close to the  calculated  data points.
Therefore, origin of the non-negligible deviation of the 
solid lines in Figs.~\ref{fig:scal-34}(c) and ~\ref{fig:scal-34}(d) 
and Fig.~\ref{fig:scal-3344}(b) from the data points is 
attributed to  $B_4^{(0)}$ given by the leading-order effective 
theory~\cite{Platter04}.
We estimate that the \mbox{theory}
underestimates $B_4^{(0)}$ by some \mbox{70--80 mK} 
($\sim\mbox{12--14}\%$ of $B_4^{(0)}$)
while it overestimates $B_4^{(1)}$ by about 2--3 mK 
($\sim\mbox{2--3}\%$ of $B_4^{(1)}$)
compared with the calculation using the realistic $^4$He potentials.

\subsection{Dimerlike-pair model}

In Fig.~\ref{fig:scal-34}(b) for the $B_3^{(0)}$-$B_4^{(1)}$ 
correlation, the dashed (blue) line, predicted by 
the dimerlike-pair model~\cite{Hiyama2012},  
is close to  the 14 data points 
with almost the same quality as the dotted (red) line
of the least squares fit.

\begin{figure}[t]
\begin{center}
\epsfig{file=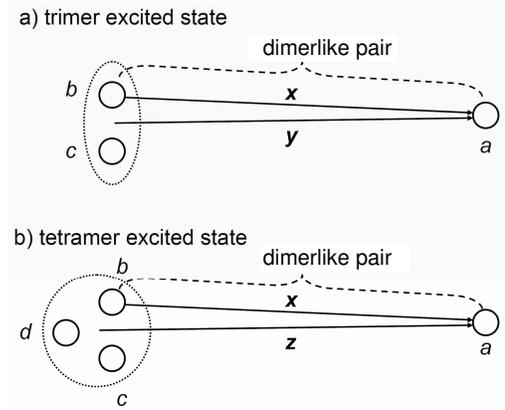,width=210pt}
\end{center}
\caption{
Schematic picture of the dimerlike-pair model 
for the trimer and tetramer excited states in the
asymptotic region. 
}
\label{fig:dimer-model}
\end{figure}

We briefly recapitulate the model.
Firstly, for the trimer excited state in Fig.~6(a),
the model indicates  that 
(i) particle $a$, located far from 
$b$ and $c$  (dimer), which are {\it loosely} bound, is little affected
by the interaction between $b$ and $c$;
(ii) therefore, the pair $a$ and $b$ at a distance $x$ 
is asymptotically dimerlike;
(iii) since \mbox{{\bf x} $\simeq$ {\bf y}}  asymptotically, 
the wave function of \mbox{particle $a$} 
along {\bf y}, exp$(-k_3^{(1)} y)/y$, is the same as 
that of the dimer, exp$(-k_2 x)/x$; hence we have  
a relation $k_3^{(1)} = k_2$.
The binding wave numbers are related to the binding energies as
\vskip -0.7cm
\begin{eqnarray}
k_2 & = & \sqrt{2\mu_x B_2}/\hbar , \nonumber \\
k_3^{(1)} & = & \sqrt{2\mu_y (B_3^{(1)} - B_2)}/\hbar , \nonumber 
\end{eqnarray}
\vskip -0.1cm
\noindent
where  $\mu_x=\frac{1}{2}m$ and $\mu_y=\frac{2}{3}m$
are the reduced mass associated with 
the coordinates {\bf x} and {\bf y}, respectively. 
Using the relation $k_3^{(1)}= k_2$, the model predicts
\vskip -0.3cm
\begin{eqnarray}
&&  \frac{B_3^{(1)}}{B_2} =   
\frac{B_2}{B_2}+\frac{3}{4}=\frac{7}{4} \quad
\Big( \frac{\Delta B_3^{(1)}}{B_2}= \frac{3}{4} \, \Big),\;\; 
\end{eqnarray}
where $\Delta B_3^{(1)}= B_3^{(1)} -B_2$ 
is the binding energy measured from the dimer.

Similarly, the model predicts  
the tetramer excited-state  energy as follows:  
Asymptotically, in Fig.~6b, 
particle $a$  decays from the trimer $(b+c+d$) as
${\rm exp}(-k_4^{(1)} z)/z$ with
\begin{eqnarray}
k_4^{(1)}  =  \sqrt{2\mu_z (B_4^{(1)} - B_3^{(0)})}/\hbar , \nonumber 
\end{eqnarray}
where  $\mu_z=\frac{3}{4}m$ is the reduced atom-trimer mass.
Using $k_4^{(1)} = k_2$ by the same reason as above, 
the model  predicts  
\begin{eqnarray}
&& \frac{B_4^{(1)}}{B_2} =   
\frac{B_3^{(0)}}{B_2}+\frac{2}{3} \quad \;\:
\Big( \frac{\Delta B_4^{(1)}}{B_2}= \frac{2}{3} \,\Big),\;\;
\quad  
\end{eqnarray}
where $\Delta B_4^{(1)}= B_4^{(1)} -B_3^{(0)}$ is the binding energy
measured from the trimer ground state.
  
We emphasize that the  relations (4.7) and (4.8) are
interaction independent.
Our assumption is only that the interparticle distance 
in the dimer (trimer)
is  larger than the interaction range, which is fulfilled 
in the present case.
The relation (4.8) is  plotted in Fig.~\ref{fig:scal-34}(b)
by the dashed (blue) line, which almost overlaps with the
dotted (red) line for the linear fit to the 14 data points by the
present few-body calculation.

\begin{table}[t]
\caption{Comparison of
the trimer (tetramer) excited-state binding energy 
$B_3^{(1)} (B_4^{(1)})$ between  
the dimerlike-pair-model prediction in the column "model" and the
present three-body (four-body) calculation.
The model predicts $\frac{\Delta B_3^{(1)}}{B_2}=\frac{3}{4}$
and $\frac{\Delta B_4^{(1)}}{B_2}=\frac{2}{3}$.  
}
\label{table:1}
\begin{center}
\begin{tabular}{ccccccc} 
\hline \hline
\noalign{\vskip 0.2 true cm} 
             & \multicolumn{2}{c} {Model}  & 
              \multicolumn{4}{c}  {Present calculation}   \\
\noalign{\vskip -0.2 true cm} 
  &    \multispan2 {\hrulefill} & 
              \multispan4 {\hrulefill} \\
  Potential &  $B_3^{(1)}$ & $\;$ $B_4^{(1)}$  &$\;$ $\;$   
$B_3^{(1)}$ & $B_4^{(1)}$   
  &  $\frac{\Delta B_3^{(1)}}{B_2}$  &  $\frac{\Delta B_4^{(1)}}{B_2}$   \\
  &  (mK) & (mK) & (mK) & (mK)  &  &  \\
\noalign{\vskip 0.05true cm} 
\hline 
\noalign{\vskip 0.2 true cm} 
LM2M2 &   2.29  &   127.37  & $\;$$\;$   2.2779  &$\;$127.42 
 & $\;$ 0.74  &  0.70 \\
TTY   &   2.30  &  127.33  &  $\;$$\;$  2.2844  &$\;$127.37  
& $\;$ 0.74  &  0.70 \\
HFD-B3-FCI1 &   2.53  &  129.96 &$\;$$\;$   2.4475 &$\;$129.89  
& $\;$ 0.69  &  0.61  \\
CCSAPT07 &  2.74  &  132.05 & $\;$$\;$  2.5890  &$\;$131.88  
& $\;$ 0.66  &  0.56  \\
PCKLJS &   2.83   &  132.91 &  $\;$$\;$   2.6502  &$\;$132.70  
& $\;$ 0.64  &  0.53  \\
HFD-B  &   2.96   &  134.21 &  $\;$$\;$   2.7420  &$\;$133.94  
&$\;$  0.62  &  0.51  \\
SAPT96 &   3.05   &  135.18 & $\;$$\;$    2.8045  & $\;$134.86  
&$\;$  0.61  &  0.48  \\
\noalign{\vskip 0.1 true cm} 
\hline
\hline
\end{tabular}
\label{table:dimerlike-model}
\end{center}
\end{table}

The trimer and tetramer excited-state binding energies 
predicted by the dimerlike-pair 
model are summerized in \mbox{Table VI} 
in comparison  with the results of the
present three- and four-body calculations  using various 
$^4$He interaction. 
Error of 
the model prediction 
is known to be  0.01--0.25 mK in $B_3^{(1)} (\Delta B_3^{(1)})$ and
0.05--0.32 mK in $B_4^{(1)} (\Delta B_4^{(1)})$ compared with
the three-(four-)body calculation. The model works well
when we compare, in Fig.~3(b) (the $B_3^{(0)}$-$B_4^{(1)}$ correlation), 
the 14 data points  with 
the dashed (blue) line and with the solid line.
 
However, strictly speaking about the model, 
it predicts \mbox{neither} $B_3^{(1)} (B_4^{(1)})$ nor the threshold
energy $B_2 (B_3^{(0)})$ but 
does predict their difference $\Delta B_3^{(1)} (\Delta B_4^{(1)})$. 
Therefore, the \mbox{error} or accuracy of the model prediction
might be judged also on the basis of  $\Delta B_3^{(1)} (\Delta B_4^{(1)})$. 
For this purpose, one can compare in Table VI the value of 
$\frac{\Delta B_3^{(1)}}{B_2}\, (\frac{\Delta B_4^{(1)}}{B_2})$
by the three-\mbox{(four-)}body calculation with the value,
$\frac{3}{4}\, (\frac{2}{3})$,  by the model.
The relative error of the model prediction for
$\frac{\Delta B_3^{(1)}}{B_2}\, (\frac{\Delta B_4^{(1)}}{B_2})$
is estimated as 1--19\% (5--28\%)\footnote{One sees in Table VI 
that the error of the model increases
with increasing $B_2$ from the top to the bottom. 
However, investigation of the
reason for this behavior is out of the scope of the present work.}.
This is not very small, but we remark that
the simple model provides, without any elaborate calculation,
a reason why the quantity
$\frac{\Delta B_3^{(1)}}{B_2}$ appears close to $\frac{3}{4}$
\mbox{(0.61--0.74 in Table VI)}, and  
$\frac{\Delta B_4^{(1)}}{B_2}$  is close to $\frac{2}{3}$
\mbox{(0.48--0.70)}.


Generally, for the $^4$He$_N$, the model suggests 
a relation 
\begin{equation}
\frac{B_N^{(1)}}{B_2} = \frac{B_{N-1}^{(0)}}{B_2} +  \frac{N}{2(N-1)} ,
\end{equation}
where the last term is ratio of
the reduced mass of the dimerlike pair ($\frac{1}{2}m$) 
to that of the \mbox{$^4$He$\,$-$^4$He$_{N-1}$ system ($\frac{N-1}{N}m$).}

\section{Summary}

Using the Gaussian expansion method for \mbox{{\it ab initio}} variational
calculations of few-body 
systems~\cite{Kamimura88,Kameyama89,Hiyama03}, 
we have calculated the binding energies of the $^4$He trimer and tetramer
ground and excited states,  
$B_3^{(0)}, B_3^{(1)}, B_4^{(0)}$ and $B_4^{(1)}$, with the use of 
the currently  most accurate  $^4$He potential
proposed by Przybytek {\it et al.}~\cite{PCKLJS},
called the PCKLJS potential.
This is an extension of our previous work~\cite{Hiyama2012}
using the LM2M2 potential.
Employing the PCKLJS, 
LM2M2, TTY, HFD-B, HFD-B3-FCI1, SAPT96 and CCSAPT07 potentials,
we have calculated the  four kinds of the binding energies 
and investigated the correlations 
between any two of them.

The main conclusions are summarized as follows:

(i) We obtained, using PCKLJS, $B_3^{(0)}=131.84$ mK,
$B_3^{(1)}=2.6502$ mK (1.03 mK below the dimer),  
$B_4^{(0)}=573.90$ mK and
$B_4^{(1)}=132.70$ mK (0.86 mK below the trimer ground state).
This potential includes
the adiabatic, relativistic, QED and residual 
retardation corrections.
Contributions of the  corrections to the
tetramer eigenenergy $-B_4^{(0)} (-B_4^{(1)})$ are, respectively, 
$-4.13\,(-1.52)$ mK, $+9.37 \,(+3.48)$ mK,  
 $-1.20 \,(-0.46)$ mK and $+0.16\,(+0.07)$ mK;
the entire correction is +4.20 (+1.57) mK.

(ii) The six correlations
\mbox{$B_3^{(1)}$-$B_4^{(1)}$}, \mbox{$B_3^{(0)}$-$B_4^{(1)}$},
\mbox{$B_3^{(1)}$-$B_4^{(0)}$}, \mbox{$B_3^{(0)}$-$B_4^{(0)}$},
\mbox{$B_3^{(0)}$-$B_3^{(1)}$} and \mbox{$B_4^{(0)}$-$B_4^{(1)}$},
are observed to be all linear 
for the various  $^4$He potentials mentioned above
over the range of 
binding energies relevant for the $^4$He atoms.
They may be called the generalized Tjon lines.
The universal scaling curves
given by the leading-order effective theory for the
$^4$He atoms~\cite{Platter04} locate
closely to the presently-obtained linear lines 
in the \mbox{$B_3^{(0)}$-$B_4^{(1)}$},
\mbox{$B_3^{(1)}$-$B_4^{(1)}$}
and \mbox{$B_3^{(0)}$-$B_3^{(1)}$} correlations, but 
deviate non-negligibly from the lines in  the 
\mbox{$B_3^{(0)}$-$B_4^{(0)}$}, \mbox{$B_3^{(1)}$-$B_4^{(0)}$}
and \mbox{$B_4^{(0)}$-$B_4^{(1)}$} correlations.
The latter deviations are attributed to the fact that
the leading-order \mbox{effective} theory underestimates $B_4^{(0)}$
by some 70--80 mK ($\sim$12--14\% of $B_4^{(0)}$) compared with the
present calculations using the realistic $^4$He potentials.

(iii) As long as the binding energies of the 
excited states of the trimer and tetramer,
the interaction-independent prediction, 
$ \frac{B_3^{(1)}}{B_2} =  \frac{7}{4}$ and
$ \frac{B_4^{(1)}}{B_2} =   
\frac{B_3^{(0)}}{B_2}+\frac{2}{3}$, 
by the dimerlike-pair model~\cite{Hiyama2012}
explains  the  \mbox{$B_2$-$B_3^{(1)}$} and  
\mbox{$B_3^{(0)}$-$B_4^{(1)}$}
correlations for the various 
$^4$He potentials, with $B_2$ being the dimer binding energy.

\section*{Acknowledgement}
We thank Krzysztof Szalewicz for 
providing us with codes producing the $^4$He 
potential (PCKLJS) used in this work.
The numerical calculations were performed on \mbox{HITACHI SR16000}
at KEK and YIFP.


\end{document}